\documentclass{elsart}

\usepackage{graphicx}

\usepackage{amssymb}

\begin{document}

\begin{frontmatter}

\title{Omori Law for Sliding of Blocks on Inclined Rough Surfaces\thanksref{support}}
\thanks[support]{Work supported in part by CNPq, FINEP and PRONEX (Brazilian government agencies).}

\author{E. J. R. Parteli}
\address
{
Institut f\"ur Computeranwendungen, Universitaet Stuttgart, Pfaffenwaldring 27, 70569, Stuttgart, Germany.
}
\author{M. A. F. Gomes\corauthref{mafg}}\corauth[mafg]{Corresponding author. fax:+55-81-3271-0359}\ead{mafg@ufpe.br}\and{\author{E. Montarroyos}}

\address
{
Departamento de F\'{\i}sica, Universidade Federal de Pernambuco, 50670 - 901, Recife, PE, Brazil.
}
\author{V. P. Brito}
\address{
Departamento de F\'{\i}sica, Universidade Federal do Piau\'{\i}, 64049 - 550 Teresina, PI, Brazil
}

\begin{abstract}
Long sequences of slidings of solid blocks on an inclined rough surface submitted to small controlled perturbations are examined and scaling relations are found for the time distribution of slidings between pairs of large events as well as after and before the largest events. These scaling laws are similar to the Omori law in seismology but the scaling exponents observed are different. Log-periodicity correction to the Omori scaling is also found. It is shown that the scaling behaviors are dependent on the angle that the incline forms with the horizontal.

\end{abstract}

\end{frontmatter}

%main text
\section{Introduction}
Recent experiments have shown that long sequences of slidings of cylindrical blocks occurring on a rough incline submitted to small controlled perturbations give origin to nontrivial scaling laws for the measured length and lifetime distributions of events [1-3]. These slidings of blocks occur as an energy dissipation process on the interface block-incline to which the elastic energy is continuously input, and as a consequence, some connection between this problem and the nonlinear dynamics of earthquakes could be hypothesizing.

The surge of interest observed in the last few years in problems involving rough surfaces is at one time the response to newly available experimental capacity [4,5] and simultaneously the embodiment and the driving force for a shift of attention towards the dynamical physics of complex phenomena typified by avalanches on sandpiles [6-8], granular flow in general [9-12], creep [13], depinning transitions [14], and friction in general [15], among many others [16].

Although the spatial scaling law observed with sliding of blocks in Ref.1 be reminiscent of the Gutenberg-Richter law for the frequency of occurrence of earthquakes [17], we can not state that the analogy between earthquakes and sliding blocks is well stablished. Moreover, the distribution of duration of slidings [2], and recent studies on effects of persistence and intermittency in these long time series [3] suggest in addition a complex structure for the distribution of sliding events along the time. Earthquakes do not occur in a purely random way and seldom occur as isolated events, but are usually part of a sequence with foreshocks and aftershocks associated with a larger event called the mainshock. Frequently the (fore)aftershocks are defined so that the largest (fore)aftershock in the sequence is typically one order of magnitude smaller than the mainshock. Sequences of earthquakes not associated with a dominant event are called swarms. The decay of aftershock sequences obeys the Omori law (for his observation of it following the 1891 Nobi earthquake) which states that the rate $n(t)$ of aftershocks at time $t$ after the mainshock decreases as a power law [17]. A similar Omori law is also reported to appear in hydraulic fracturing [18]. This paper reports results on the time distribution of mechanical slidings of blocks on an inclined rough surface which are reminiscent of the Omori law, although with a scaling exponent different from that observed for real earthquakes. 

\section{Experimental details}

The time series of mechanical sliding lengths \{${\lambda}(t)$\} studied in this paper were obtained with the fully automatic 2000 mm long V-shaped aluminum incline with aluminum blocks used in Ref.3. This incline can be fixed at any angle $\theta$ with the horizontal with a precision of ${0.02}^{\circ}$. The apparatus is held fixed on a table which is isolated from mechanical vibrations and temperature and humidity are constant. A laser assisted optical device measures the sliding lengths ${\lambda}(t)$ with a precision of $0.5{\, \mbox{mm}}$ and the data is stored and analyzed in a microcomputer. The discrete time $t=1,2,{\ldots},T$ here is defined by impacts of an electronic hammer which provides small controlled perturbations on the incline. Twenty four series of slidings were studied in all, each one corresponding to a different angle of the incline. The total number of impacts in these series varied from $T$=500 to $T$=5000 and the number of impacts needed to move the solid block along a single full length of the incline varied from 182 to 700. When the cylinder collides with the end of the groove, the last sliding is neglected and the cylinder is automatically placed on the top of the incline and a new chain of events initiates. For each one of these sequences, we chose the events with much larger magnitude than the average sliding length ${\bar{\lambda}}$, and we identified them as the mainslidings (MS) of the series. Some pairs of MS were selected and the total number of slidings $N(t)$ between the nearest neighbors MS were recorded as a function of $t$, with $t$=0 associated to the first MS. Afterwards, events between MS with magnitude $\lambda < q{\cdot}{\lambda}_{\mathrm{MS}}$, were also recorded for different values of $q$ between 0.005 and 1.0. Sequences of swarm-like slidings although occurring in the experiments will not be studied in this work.

\begin{figure}
\begin{center}
\includegraphics*[width=0.95\columnwidth]{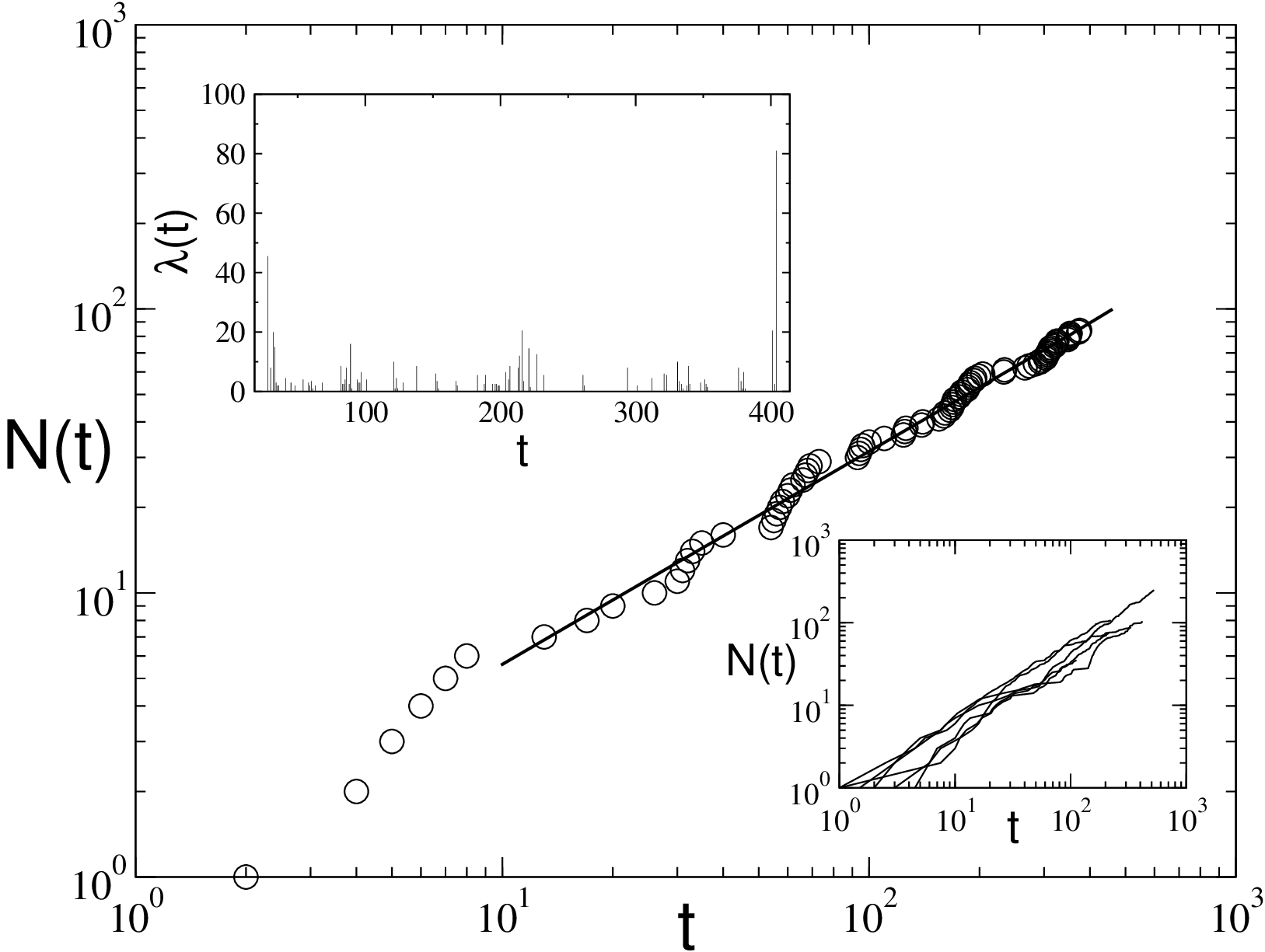}
\caption{The central plot shows in log-log coordinates the number $N(t)$ of slidings observed between two mainslidings (MS) as a function of the number $t$ of impacts on the incline, in the domain of small angles in a single experiment ($\theta = {18.0}^{\circ}$ in this case, and ${\theta}_{\mathrm{c}}={32.0}^{\circ} $). The sliding lengths $\lambda$ are given in mm, and the length(diameter) of the aluminum block on the incline was 500\,mm(12\,mm). The variable $t$ at the first MS, and the filter factor $q$ have the values $0$ and $1$, respectively. The continuous line represents the best fit $N \, {\sim} \, t^{\, 0.74}$. The corresponding time series is shown in the upper-left-hand-corner inset. The bottom-right-hand-corner inset shows the time evolution of $N(t)$ for six experiments in the domain of large inclinations. In this last situation $N\, {\sim} \, t^{\, 1}$. See text for details.}
\label{fig:fig1}
\end{center}
\end{figure}

If the angle $\theta$ which the incline forms with the horizontal is close to ${\theta}_c = {\tan}^{-1} {\mu}_s$ \ (${\mu}_s$ = coefficient of static friction) in the interval $0.01<({\theta}_c - {\theta})/{\theta}_c<0.25$, \ $N(t)$ is observed to increase trivially as $N(t)=k{\cdot}t$, with $0.5<k<1$, as shown in the bottom-right-hand-corner inset of Fig.1 with reduction factor $q=1.0$ for 6 typical sliding series. In this case the rate $n(t)=dN(t)/dt$ \ is simply the constant $k$. A similar behavior was observed for all the other values of $q$.

On the other hand, for smaller angles in the interval $0.20<({\theta}_{c}-{\theta})/{{\theta}_{c}}<0.45$, the sliding response of the block-incline system becomes complex and a large and variable number of mechanical perturbations (and consequently of time units) is needed before each sliding occurs; i.e. the system in this domain of angles is much more slowly driven than in the phase of large angles. Eight series in this domain of small angles were studied and a typical sequence of slidings between two MS ($\lambda$=46\,mm at $t$=0, and $\lambda$=81\,mm at $t$=376) is shown in the upper-left-hand-corner inset of Fig.1. The reader can observe in this figure that there is intermittency (regions with bursts of sliding activity separated by regions without activity) in the distribution of slidings; i.e. slidings are no longer uniformly distributed as occurs for large angles. A detailed examination in the sequence of slidings of Fig.1 shows that in 77$\%$ of the total time span of this sequence the block does not move ($\lambda$=0), and in the remaining 23$\%$ there are 86 slidings. The average sliding length between the two MS in this particular case is ${\bar{\lambda}}={\sum_{i=1}^{375} {\lambda}_i/375}$ = 1.35\,mm; thus the MS are much larger than $10\,{\bar{\lambda}}$ for this sequence $-$ a similar proportion is valid for all sequences of slidings between MS studied in this paper. Another interesting aspect of this inset lies in a fact that is commonly observed in real earthquake sequences: a high concentration of aftershocks following the MS and a low concentration of foreshocks before a MS. 

The central plot of Fig.1 exhibits $N(t)$ (with $q=1.0$) for the sequence shown in the inset; the solid line represents the best fit $N(t)=t^{\, 0.74}, \ \ 10<t<380$. The reader may observe that the main plot in Fig.1 shows some discontinuities which are connected to the regions where the bursts of intermittency occur. This scaling behavior is independent of $q$. The main plot of Fig.2 exhibits the total number of afterslidings ${N_{\mathrm{as}}(t)}$ associated to the first MS of Fig.1 for $q=0.1$. The corresponding time series is shown in the inset of Fig.2. These slidings are restricted to the first half of the time interval shown in the inset of Fig.1, and satisfy $\lambda < {0.1}\,{\lambda}_{\mathrm{MS}}$. The best fit given by the continuous line in Fig.2 has a slope of 0.73, which is essentially the same slope observed in Fig.1 within the typical experimental uncertainties of \ 5 - 10$\%$. The scaling exponent for aftershocks does not depend on the filter factor $q$ within the error bars. Furthermore, if we focus our attention in the foreslidings of the second MS appearing in Fig.1 (i.e. in events distributed on the second half of the series shown in the inset of Fig.1), a similar scaling relation can again be obtained although with a higher level of statistical fluctuations due to the relative reduction in the number of events which characterizes this part of the sequence preceding a MS. 

\begin{figure}
\begin{center}
\includegraphics*[width=0.80\columnwidth]{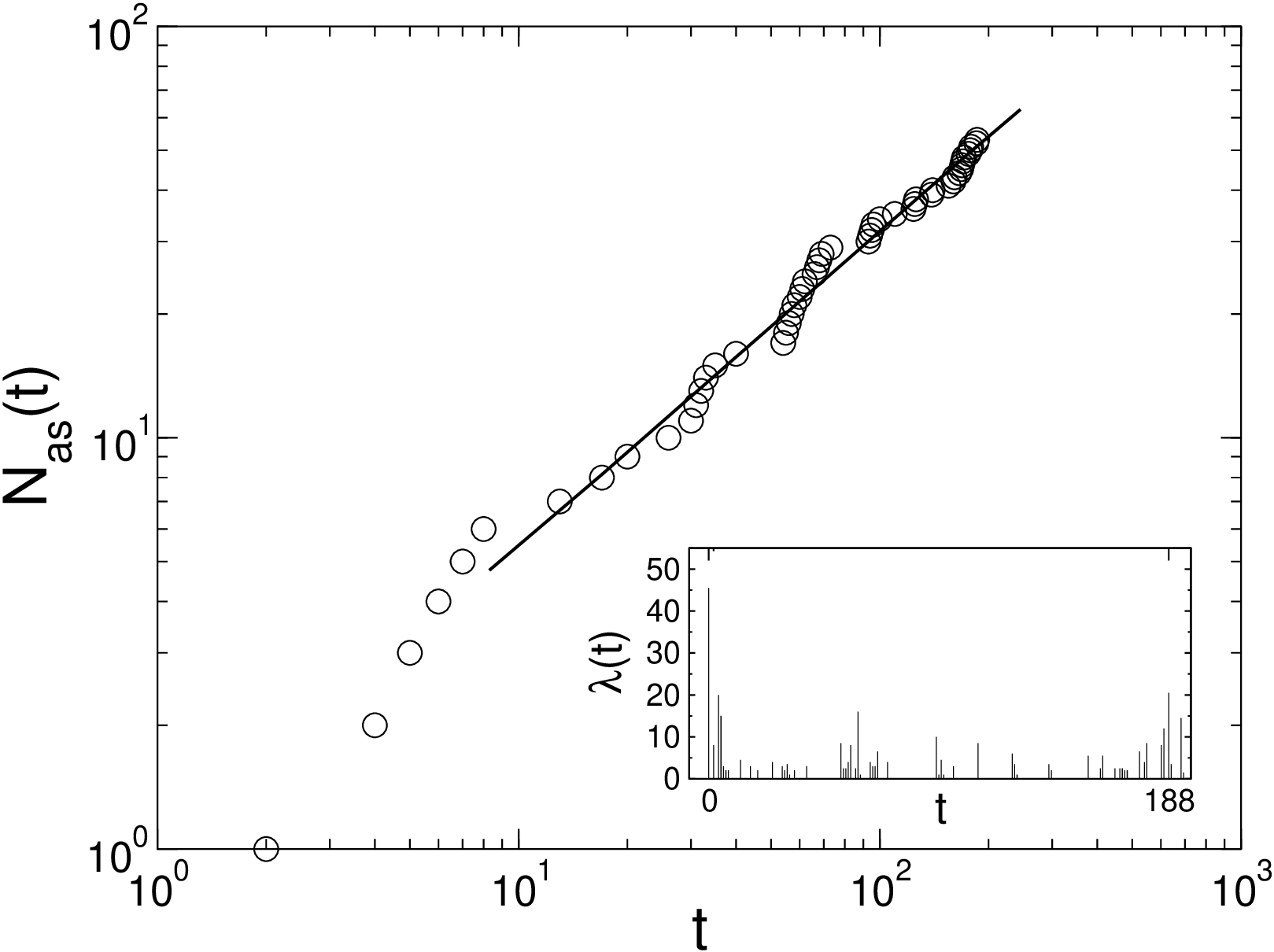}
\caption{The same as in the main plot of Fig.1 but restricted to the aftershocks of the first MS. The filter factor $q$ is $0.1$, and the continuous line has the slope 0.73. See text for details.}
\label{fig:fig2}
\end{center}
\end{figure}

\begin{figure}
\begin{center}
\includegraphics*[width=0.60\columnwidth]{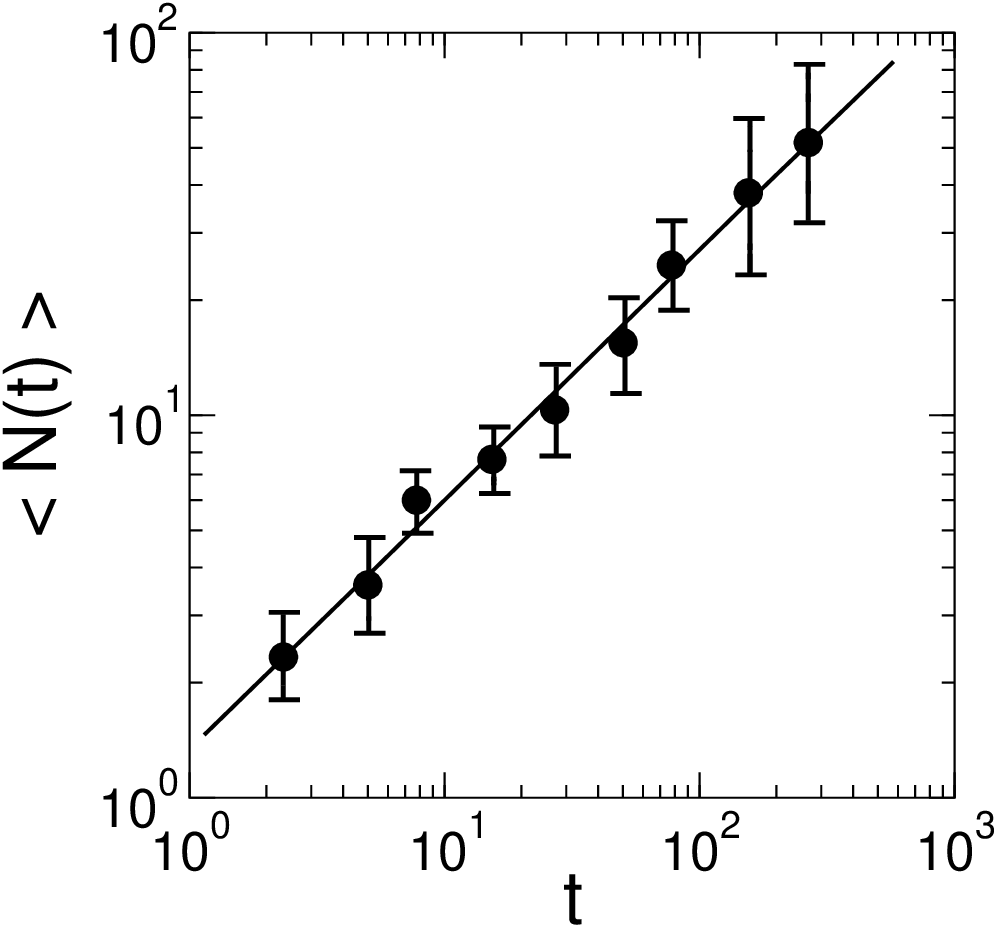}
\caption{The average number of slidings $\left< N(t) \right>$ between nearest neighbors MS as a function of the time for all the experiments in the domain of small angles and for $q=1.0$. The vertical bars give the statistical fluctuations. The straight line has the slope $0.65 \, \pm \, 0.10$.}
\label{fig:fig3}
\end{center}
\end{figure}

\begin{figure}
\begin{center}
\includegraphics*[width=0.9\columnwidth]{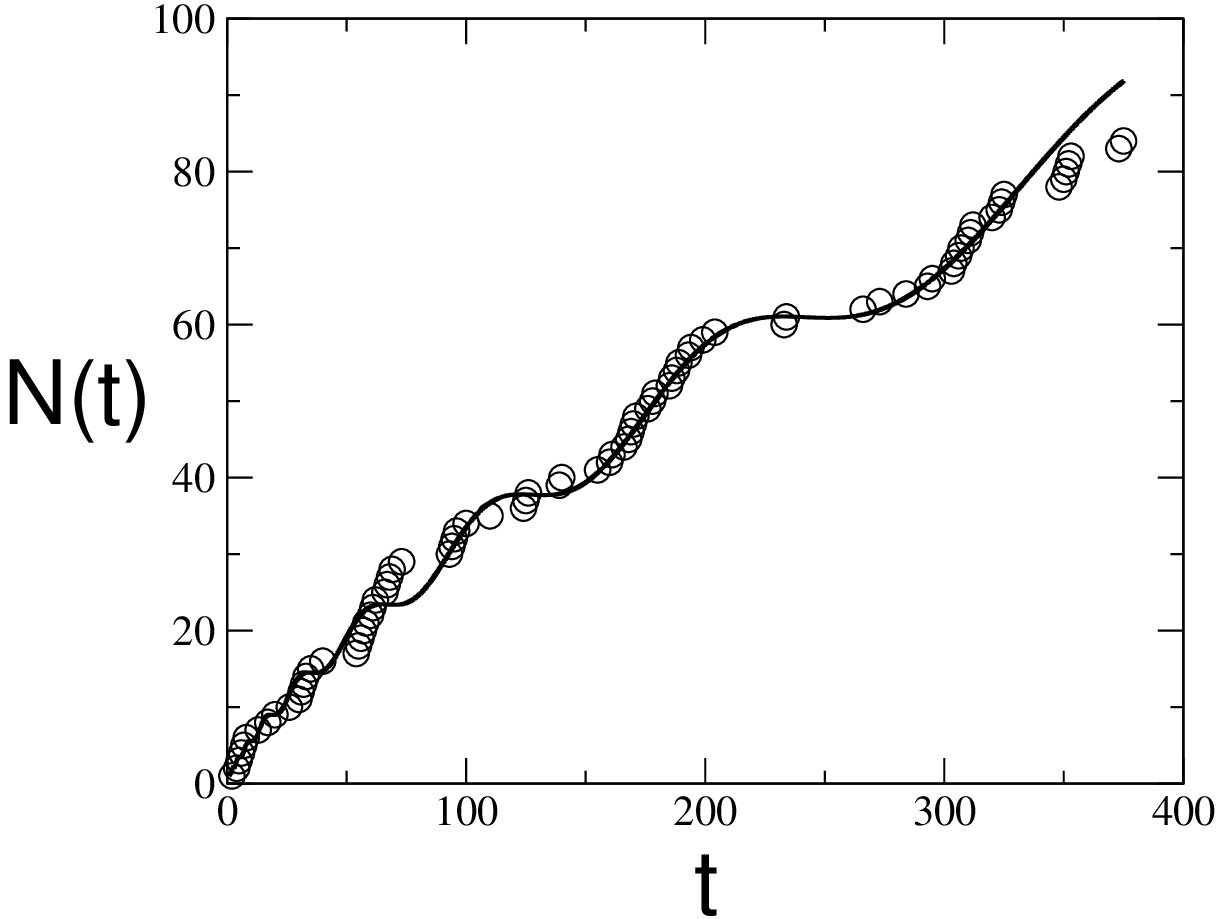}
\caption{The observed number of slidings, $N(t)$, for the time series of Fig.1 together with the adjust given by Eq.3 (continuous line): \ $t^{\, 0.75} \left[1+0.082\,{\cos{({9.84}\ln{t}-{2.09})}} \right]$.}
\label{fig:fig4}
\end{center}
\end{figure}

Different sequences of slidings give essentially the same exponent independently of the magnitude of the MS. In order to see how this works, we show in Fig.3 the average number of slidings between two MS, $\left< N(t) \right>$, as a function of the time from the first MS, and the respective statistical fluctuations for all sequences of slidings studied. The solid line represents the best fit

\begin{equation}
\left< N(t) \right> \ {\sim} \ t^{\, (1-p)}, \ \ \ \ \ \ t>1, \ \ \ \ p=0.35,
\end{equation}

\noindent
along two decades of variability in time. Our overall estimate for the exponent $p$ is $0.35 \, \pm \, 0.10$, as indicated by the fluctuation bars in Fig.3. Within the margins of errors, no change of the scaling exponent was observed after variations of the value of $q$, i.e. the power law in (1) is very robust irrespective of the filter factor. These results lead to a decay rate in the average number of slidings of blocks between pairs of MS as

\begin{equation}
\left< n(t) \right> {\sim} \, t^{-p}=t^{-(0.35 \, \pm \, 0.10)}, \ \ \ \ \ \ t>1,
\end{equation}

\noindent
an expression reminiscent of the Omori law for real earthquakes, although with a sensibly smaller exponent. The standard Omori exponent is $p_{\mathrm{Omori}} \simeq 1.0$ [17]. However, observations during the last decades [19$-$22] have pointed toward a generalized Omori law in the form $n(t)=K(c+t)^{-p}$, with $K$ and $c$ constants, and $0.6<p<2.0$, as an appropriate representation of the temporal variation of aftershock activity. The exponent $p$ in these cases differs from sequence to sequence, and it is presently not clear which factor is most significant in controlling this quantity. Shaw [23], in particular, derived the generalized Omori law with scaling exponents close to one using a theory that involves the response to sudden forcing of a dynamics of self-driven acceleration to failure.

Recent advance in the theory of fracture and fragmentation has been invoked to explain empirical laws in seismology [24]. In this context, it is interesting to notice that Eq.1 has an analogy in fragmentation dynamics, namely the scaling relation $N_{\mathrm{frag}}(t) \, {\sim} \, t^w$ giving the (average) total number of fragments produced by the dynamics from some main fragment existing originally at $t$=0. Computer simulations [25] in these cases indicate that $w=0.80 \, \pm \,  0.10$ for 1d fragmentation controlled by diffusive processes. The numerical resemblance of the exponent $w$ with the exponent $1-p$ (see Eq.1) and the fact that the rough interface cylinder-incline in our experiments is essentially a 1d-system, as well as the probable importance of diffusive processes to explain the emergence of both the Gutenberg-Richter law and the Omori law [26], suggest that the picture of sliding activity between MS as a fragmentation dynamics acting in the space of contacts at the interface block-incline could be a good one. 

A new way to improve the accuracy of the description of irreversible and intermittent failure processes associated with time series of real earthquakes introduces log-periodic corrections to the leading scaling behavior [27]. Inspired in this approach we tried to adjust our experimental time series of slidings with the scaling form

\begin{equation}
N(t)=t^{1-p}\left[1+b\,{\cos{({\omega}\ln{t}-{\varphi})}}\right].
\end{equation}

\noindent
In Eq.3, the correction \ $t^{\, 1-p}\left[b\,{\cos{({\omega}\ln{t}-{\varphi})}}\right]$ to the leading scaling behavior $t^{\, 1-p}$ of the Omori law is essentially the real part of the power law $t^{\, z}$, whose scaling exponent is the complex number \ $z=(1-p)+{\mbox{i}}{\omega}$, with the imaginary part of $z$ corresponding to the angular frequency $\omega$. Novikov in 1969 [28] pointed out the possibility of log-periodic corrections to explain experimental results on intermittency involving fluids, and Nauenberg in 1975 [29] studied a scaling representation containing a log-periodic dependence in the singular part of the free energy of systems described by scaling operators which exhibit a critical phase transition. In Fig.4, we show in a linear plot the experimental data of Fig.1 adjusted with the four parameter expression given in (3), with $p=0.25$, $b=0.082$, ${\omega}=9.84$, and ${\varphi}=2.09$. All time series of slidings studied in this work are equally quite well described by Eq.3 with similar fitting parameters; in particular, the typical variation in the exponent $p$ is \ $0.25<p<0.36$; \ $b$ varies in the interval $0.075$ to $0.17$; \ $7<{\omega}<10$, and \ $0<{\varphi}<2.1$. \ Real sequences of earthquakes [27] and time series of stock exchange indexes [30] present commonly angular frequency $\omega$ (exponent $p$) in the intervals \ $2<{\omega}<15$ \ ($0.50<p<0.75$), and \ $5<{\omega}<8$ \ ($0.40<p<0.88$), respectively. The time series examined here include a large number of events comparatively to the time series of earthquakes studied in Ref.27. We believe that Eq.3 represents a true log-periodic correction to the Omori scaling in our time series of slidings and not a spurious log-periodicity generated by limitations in the sampling procedure [31].

Finally, we would like to point out that recently, Lima et al [32] using a simple model in which the coefficient of friction is a random function of the position, obtained the same static exponent found for the experimental size distribution of slidings of Ref.1. However, the average sliding length in Ref.32 scales as ${\bar{\lambda}} \sim {({\theta}_c - {\theta})}^{-1}$, while the experimental result is ${\bar{\lambda}}_{\mathrm{exp}} \sim {({\theta}_c - {\theta})}^{-{\beta}}$, with $0.14 \leq \beta \leq 0.32$ [1]. In view of the results reported in the present work, it would be highly interesting to verify if the model of Ref.32 can reproduce the dynamic scaling with log-periodic oscillations given in Eq.3, as well as to compare the behavior of this model with sequences of real earthquakes. 

\section{Conclusions}

We have shown from experiments that sequences of slidings of a solid block on a rough incline submitted to small controlled perturbations have a complex time structure, corroborating recent experimental results reported in [3]. This situation could have an analogous counterpart in seismology since most models of earthquakes reproduce easily the spatial behavior of this phenomenon which is associated with the Gutenberg-Richter law [17], but not the Omori law for aftershocks [17,19,26]. In this work, we have reported results that indicate that sequences of slidings between mainslidings as well as sequences of slidings after(before) a mainsliding obey the Omori scaling law with an anomalous and robust exponent provided the angle which the incline forms with the horizontal is not too large. In this situation the time series of slidings necessarily presents intermittency [3]. If, on the contrary, the inclination increases, the anomalous Omori scaling disappears together with the intermittent behavior. Log-periodic correction to this anomalous Omori scaling seems important to a full description of the time behavior of these sequences of slidings. 

\begin{ack}
We are grateful to G.L.Vasconcelos for a critical reading in the manuscript. E. J. R. Parteli acknowledges a fellowship from CAPES - Bras\'{\i}lia/Brasil.

\end{ack}

\end{document}